\documentclass[twocolumn]{aastex62}
\usepackage[section]{placeins}
\usepackage{longtable}
\usepackage{hyperref}
\usepackage{url}
\usepackage{fontawesome}
\usepackage{xspace}
\usepackage{comment}
\usepackage[flushmargin]{footmisc}
\graphicspath{{./}{figures/}}
\newcommand{\nancy}{\textit{Roman}\xspace}
\newcommand{\pandeia}{{\tt Pandeia}\xspace}
\usepackage{amsmath}

\newcommand{\STScI}{\affiliation{Space Telescope Science Institute, 3700 San Martin Dr, Baltimore, MD 21218, USA}}

\newcommand{\Princeton}{\affiliation{Department of Astrophysical Sciences, Princeton University, Princeton, NJ 08540, USA}}

\shorttitle{STIPS}
\shortauthors{Gomez et al.}

\begin{document}
\title{STIPS: The Nancy Grace Roman Space Telescope Imaging Product Simulator}

\correspondingauthor{Sebastian Gomez}
\email{sgomez@stsci.edu,ehan@stsci.edu}
	
\author{STIPS Development Team}
\STScI

\author[0000-0001-6395-6702]{Sebastian Gomez}
\STScI

\author[0000-0003-3858-637X]{Andrea Bellini}
\STScI

\author{Hanna Al-Kowsi}
\STScI

\author[0000-0001-6905-1859]{Tyler Desjardins}
\STScI

\author[0000-0003-1509-9966]{Robel Geda}
\Princeton

\author[0000-0001-9797-0019]{Eunkyu Han}
\STScI

\author[0000-0002-4679-5692]{O. Justin Otor}
\STScI

\author[0000-0003-1645-8596]{Adric Riedel}
\STScI

\author[0000-0003-0894-1588]{Russell Ryan}
\STScI

\author{Isaac Spitzer}
\STScI

\author{Brian York}
\STScI

\begin{abstract}
The Space Telescope Imaging Product Simulator (STIPS) is a Python-based package that can be used to simulate scenes from the upcoming \textit{Nancy Grace Roman Space Telescope} (\nancy). STIPS is able to generate post-pipeline astronomical images of any number of sensor chip assembly (SCA) detectors, up to the entire 18-SCA Wide-Field Instrument array on \nancy. STIPS can inject either point spread functions generated with {\tt WebbPSF}, or extended sources in any of the \nancy filters. The output images can include flat field, dark current, and cosmic ray residuals. Additionally, STIPS includes an estimate of Poisson and readout noise, as well as an estimate of the zodiacal background and internal background from the telescope. However, STIPS does not include instrument saturation, non-linearity, or distortion effects. STIPS is provided as an open source repository on GitHub \href{https://github.com/spacetelescope/STScI-STIPS}{\faGithub}.
\end{abstract}

\keywords{methods: observational -- techniques: photometric -- telescopes}

\section{Introduction}\label{sec:intro}

Originally, the Space Telescope Imaging Product Simulator (STIPS) was developed as a Python-based tool to simulate images from the telescopes managed by the Space Telescope Science Institute (STScI), namely the \textit{James Webb Space Telescope (JWST)}, \textit{Hubble Space Telescope (HST)}, and \textit{Nancy Grace Roman Space Telescope} (\nancy)\footnote{\ Formerly known as WFIRST.}. Since then, the development of STIPS has focused exclusively on simulating products from \nancy.

\nancy is a wide-field infrared telescope prepared for launch as soon as October 2026, and no later than May 2027 \citep{Spergel15}. The imager onboard the telescope will be able to observe in 8 filters covering the $\sim 0.48 - 2.3\ \mu$m range down to $\sim 24 - 26$ mag limits in $1$ minute of exposure time, depending on the filter. The Wide-Field Instrument (WFI) on board \nancy is composed of 18 near-infrared-sensitive Teledyne H4RG-10 sensor chip assemblies (SCAs) capable of covering a total field of view of $\sim 0.28$ sq. deg. with a resolution of $\sim 0\farcs11$ per pixel. A more detailed description of the instrument parameters can be found in the \nancy Science and Technical Overview\footnote{\ \url{https://www.stsci.edu/files/live/sites/www/files/home/roman/_documents/roman-science-and-technical-overview.pdf}}.

\begin{figure*}
    \begin{center}
        \includegraphics[width=0.9\textwidth]{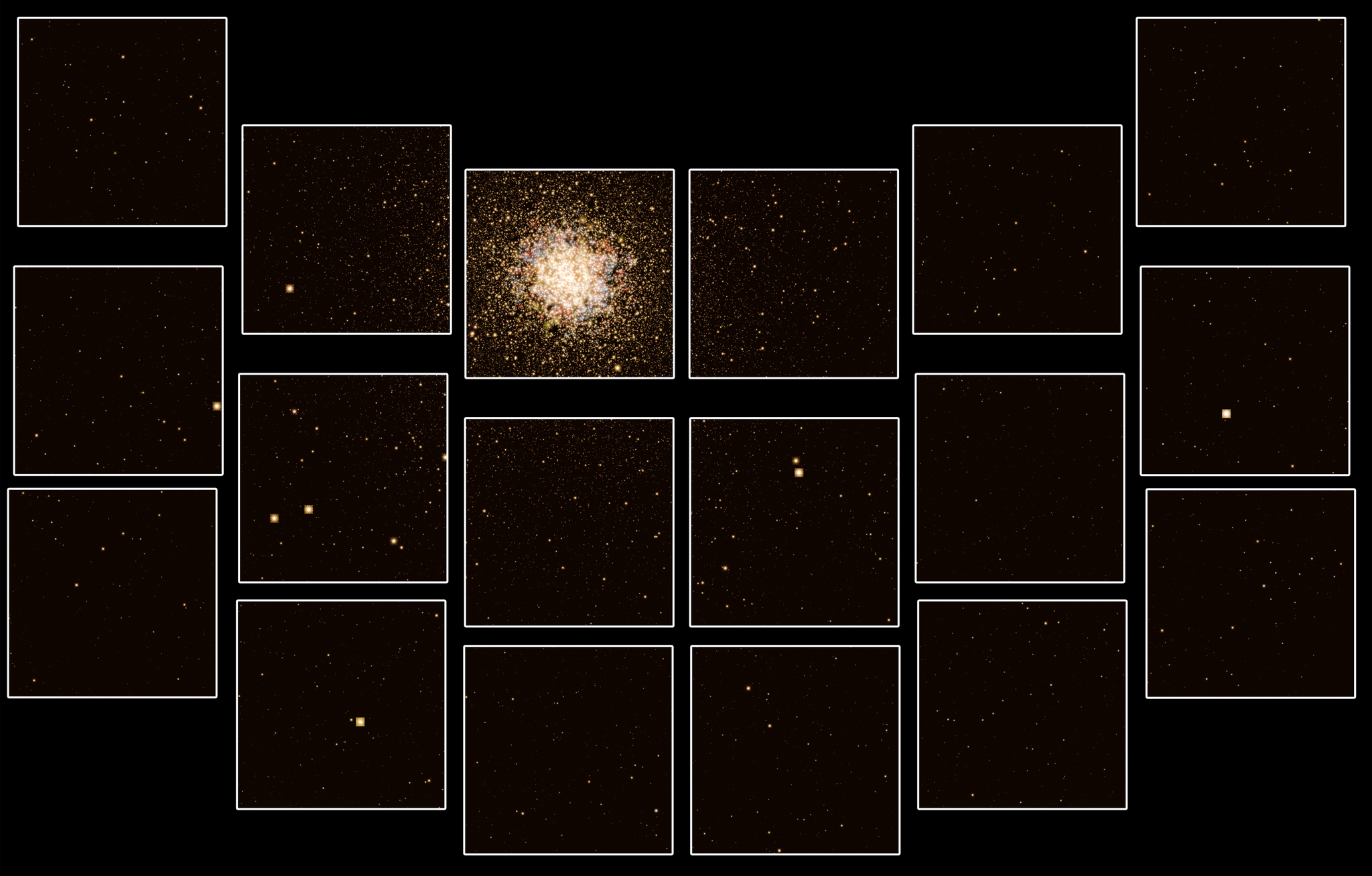}
        \caption{Simulated image of the M13 globular cluster generated with STIPS for all 18 SCAs in the WFI. The color image was created using photo editing software by combining the output {\tt fits} files from three STIPS simulations in F106, F129, and F158 filters. The magnitudes and locations of the sources in the scene were obtained from the PS1/$3\pi$ catalog \citep{Chambers16}.
        \label{fig:stips}}
    \end{center}
\end{figure*}

In this paper we describe the methods used to simulate \nancy WFI scenes with STIPS. A scene is an ideal spatial brightness distribution generated without simulating the entire light path of a telescope; but instead a representation of the post-pipeline processed image products. STIPS can inject both point sources and extended sources onto astronomical scenes. The output images can also include the effects of Poisson and readout noise, flat, dark, and cosmic ray residuals, as well as zodiacal and internal background. The point spread functions (PSFs) that STIPS injects into scenes are generated using the {\tt WebbPSF} package \citep{Perrin12}. Extended sources are generated using either \texttt{Astropy}'s \texttt{Sersic2D} or the Sérsic profile generator in \pandeia \citep{Pontoppidan16}. Estimates of the zodiacal and internal background can be directly imported from \pandeia. The zodiacal background estimates are based on the background models from JWST \citep{Rigby23}, and the internal background estimates are based on engineering data provided by the Goddard Space Flight Center (GSFC; \citealt{Ryan23}). STIPS does not include instrument saturation, non-linearity effects, or distortion effects.

STIPS is not meant to be used as a high-accuracy exposure time calculator. For that purpose, \pandeia is the recommended tool, which is designed to derive accurate signal-to-noise values for small fields of view with few sources \citep{Pontoppidan16}. STIPS is designed for rapid simulations of large fields of view; an example is shown in Figure~\ref{fig:stips}. STIPS can also be used to estimate the shape of the \nancy PSF at any location within the WFI, a feature not currently available in \pandeia. Other tools able to simulate more complex extended sources exist, including the \nancy\ module within {\tt GalSim} \citep{Rowe15}, as well as the {\tt GalSim}-based image simulator {\tt romanisim} currently under development at STScI\footnote{\url{https://romanisim.readthedocs.io/en/latest/}}, which has not yet been released and validated. The main advantage of STIPS over these last two simulators is speed. While {\tt romanisim} performs more precise simulations that include up-the-ramp fitting of individual exposures, STIPS creates idealized scenes. As a consequence, STIPS is $\sim 20$ times faster than these simulators.

In \S\ref{sec:implementation} we describe the backend of STIPS and outline the functions used to generate scenes. In \S\ref{sec:validation} we include validation metrics and estimates for the accuracy of the STIPS products. Finally in \S\ref{sec:usage} we provide references to the latest documentation and instructions to install and run STIPS.

\section{Implementation}\label{sec:implementation}

\subsection{Creating a PSF}\label{sec:PSFs}

\begin{figure*}
    \begin{center}
        \includegraphics[width=0.85\textwidth]{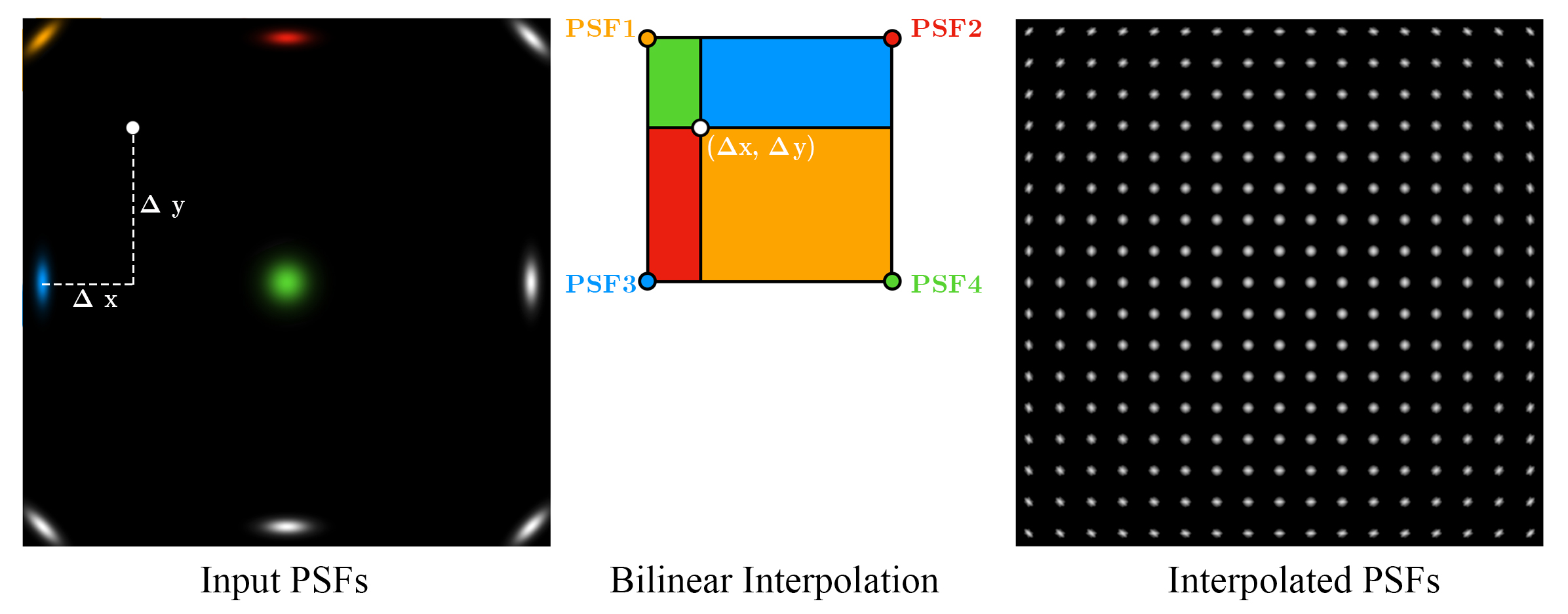}
        \caption{Graphical representation of the bilinear interpolation process that STIPS uses to estimate the shape of a PSF within a single SCA. The distortion of the PSFs shown here are greatly exaggerated for visualization purposes. \textit{Left}: Grid of nine input PSFs evenly distributed across an SCA. \textit{Middle}: To estimate the shape of a PSF at a requested location (x, y), STIPS weights the shape of the four closest PSFs by the partial area diagonally opposite to each PSF. \textit{Right}: Example PSFs distributed across the SCA generated using this method by interpolating the input PSFs on the left. \label{fig:bilinear}}
    \end{center}
\end{figure*}

The goal of STIPS is to create an accurate representation of the \nancy effective point spread function (ePSF), which is a model of the convolution of the instrumental point spread function (PSF) with the pixel-response function of the detector, integrated over individual pixels \citep{Anderson00}. Real \nancy observations will observe the ePSF, the shape of which depends on both the choice of filter and pixel position within the WFI.

We use the {\tt WebbPSF} package to create instrumental PSFs. {\tt WebbPSF} uses the latest mission-specific reference data to simulate these PSFs by interpolating a set of wavelength-dependent Noll Zernike coefficients based on GSFC optical models at five field points, sampled at the center and near the four corners, in each SCA to calculate the total wavefront at a specific position, which is then used to calculate the PSF. Given that this process can be computationally expensive, instead of computing a PSF at the pixel position of each source being simulated, STIPS interpolates a grid of nine PSFs generated with {\tt WebbPSF} with a $4\times$ oversampling uniformly distributed across the SCA being simulated. A graphical representation of the location of these nine PSFs within an SCA, the interpolation process that STIPS uses, and the output interpolated PSFs is shown in Figure~\ref{fig:bilinear}. Note that the distortion of these PSF is greatly exaggerated in that figure for visualization purposes.

In the current implementation of STIPS, the upscaling factor is fixed to a value of 4. We convolve these upscaled PSF models with the pixel-response function of the WFI, which only includes the contribution from interpixel capacitance (IPC) to create an ePSF. IPC is the phenomenon where charge collected in a pixel affects the signal in neighboring pixels, leading to signal leakage. To account for this effect, we convolve the PSF with the $3 \times 3$ IPC kernel matrix shown in Eq.~\ref{eq:ipc}, provided by the NASA GSFC on 15 Aug 2019.

\begin{equation}
{\rm IPC} = \frac{1}{100} \times 
\begin{bmatrix}
  0.21 &  1.62 & 0.20 \\
  1.88 & 91.59 & 1.87 \\
  0.21 &  1.66 & 0.22 \\
\end{bmatrix}
\label{eq:ipc}
\end{equation}

Then, depending on the pixel location of the source being simulated, STIPS will identify the four nearest ePSFs to that location. These four ePSFs are then subjected to a bilinear interpolation procedure that approximates the shape of the ePSF at the requested location. The shape of the four closest ePSFs is weighted by the partial area diagonally opposite to each PSF. This process is illustrated in the middle panel of Figure~\ref{fig:bilinear}.

\subsection{Injecting a PSF}\label{sec:injection}

After the shape of the ePSF has been determined at the requested pixel position using the procedures outlined in \S\ref{sec:PSFs}, this can be injected onto the image. The ePSF is projected into the pixel array of the SCA to calculate the corresponding brightness of each pixel. STIPS can inject the ePSF at float values X, Y that indicate the pixel position of the source on an image of size N$\times$M, which has individual pixels $p_{i,j}$, where $i$ goes from 1 to N along the $x$ direction and $j$ goes from 1 to M along the $y$ direction. The distance from the specified X, Y position to each pixel in the image is calculated using Eq.~\ref{eq:delta}, where the $\times 4$ is due to the oversampling factor of the ePSF, and $C_x, C_y$ are the pixel coordinates of the center of the ePSF.

\begin{equation}
\begin{split}
\delta_x = (i - X) \times 4 + C_x \\
\delta_y = (j - Y) \times 4 + C_y
\end{split}
\label{eq:delta}
\end{equation}

The phase $f$ of each pixel is the decimal value of the distance between the center of the ePSF and the pixel in question, calculated using Eq.~\ref{eq:phase}. The injected ePSFs are not projected onto the entire SCA, but instead only onto a smaller square region, equal to one quarter the size of the ePSF. The size of the region onto which the ePSF is projected depends on the magnitude of the source. For sources dimmer than the default of 14th mag, the values of M and N are set to 44 pixels. For sources brighter than this magnitude the range is increased to 88 pixels. For the brightest sources above the default of 4th mag, the limit is set to 176 pixels. We caution that for these brightest stars, the pixel limit does not accurately reproduce the effects of large diffraction spikes, which can extend well beyond 176 pixels. This is mostly a cosmetic effect, given that $\sim 97$\% of the point source flux remains within this box.

\begin{equation}
\begin{split}
x_{\rm int} &= {\rm int}(\delta_x) \\
y_{\rm int} &= {\rm int}(\delta_y) \\
f_x &= \delta_x - x_{\rm int} \\
f_y &= \delta_y - y_{\rm int}
\end{split}
\label{eq:phase}
\end{equation}

\begin{figure}
    \begin{center}
        \includegraphics[width=0.9\columnwidth]{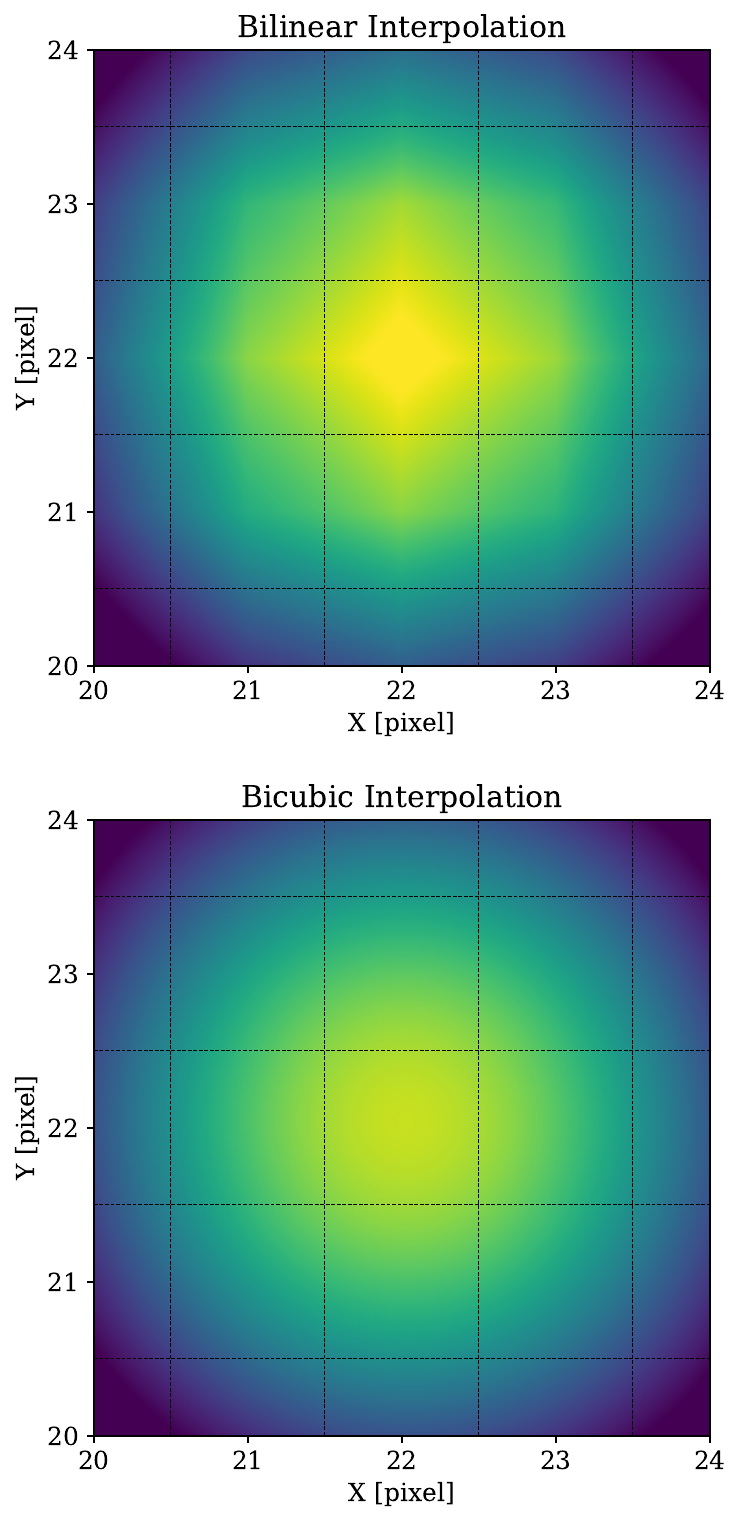}
        \caption{Input F129 \nancy ePSF displayed using a bilinear (\textit{top}) and bicubic (\textit{bottom}) interpolation method. The difference between the two methods is greatest near the code of the ePSF. The midpoint of the ePSF is at the center of pixel 22, half of the ePSF size of ${\rm N} = {\rm M} = 44$ pixels. \label{fig:interpolation}}
    \end{center}
\end{figure}

The distance between the center of the ePSF and the pixel $p_{i,j}$ is $d = \sqrt{(i - X)^2 + (j - Y)^2}$. For the pixels near the core of the ePSF with $d \le 4$, a bicubic interpolation method is used to estimate the flux of each pixel. For pixels with $d > 4$, where there is less variation from pixel to pixel, a simpler and faster bilinear interpolation method is used instead. The flux of most pixels will be estimated using a bilinear interpolation method. In Figure~\ref{fig:interpolation} we show a visual comparison between the bilinear and bicubic interpolation methods. For large values of $d$, the difference between the bilinear and bicubic interpolation methods is minimal. For small values of $d$, close to the center of the ePSF, the difference becomes more significant. The upper panel of Figure~\ref{fig:interpolation} shows how the bilinear interpolation method produces sharp artifacts near the center of the ePSF.

\begin{figure}
    \begin{center}
        \includegraphics[width=\columnwidth]{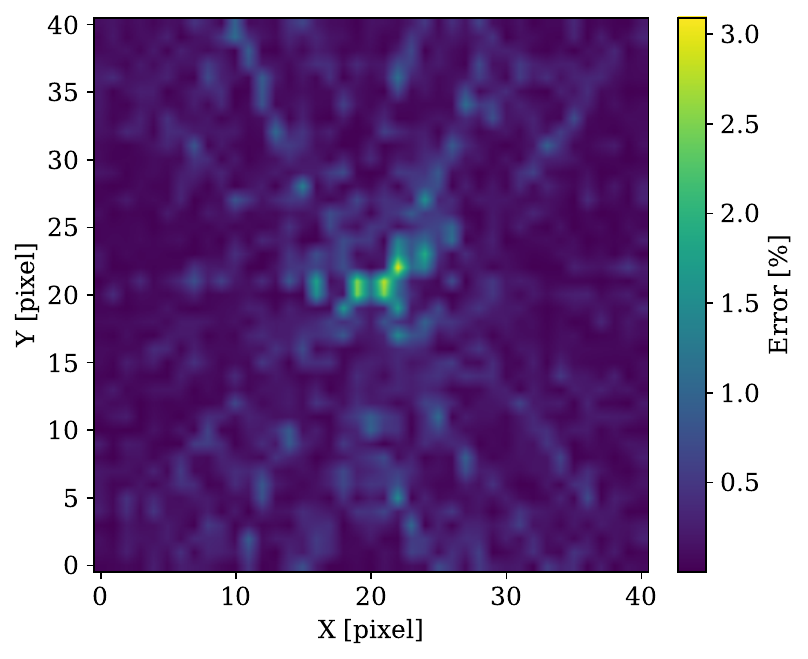}
        \caption{Error resulting from using a bilinear as opposed to a bicubic interpolation method when injecting an ePSF onto an image. The two methods are generally equivalent at the level of $\sim 1$\%, except in the core of the ePSF, where the difference is closer to $\sim 3$\%. \label{fig:interpolation_compare}}
    \end{center}
\end{figure}

In Figure~\ref{fig:interpolation_compare} we show a quantitative difference between the bilinear and bicubic interpolation methods. For $\sim 98$\% of pixels the difference between the two methods is $< 1$\%. In the core pixels with $d < 4$, the error resulting from using the bilinear interpolation method increases to $\sim 3$\%.

\begin{figure*}
    \begin{center}
        \includegraphics[width=\textwidth]{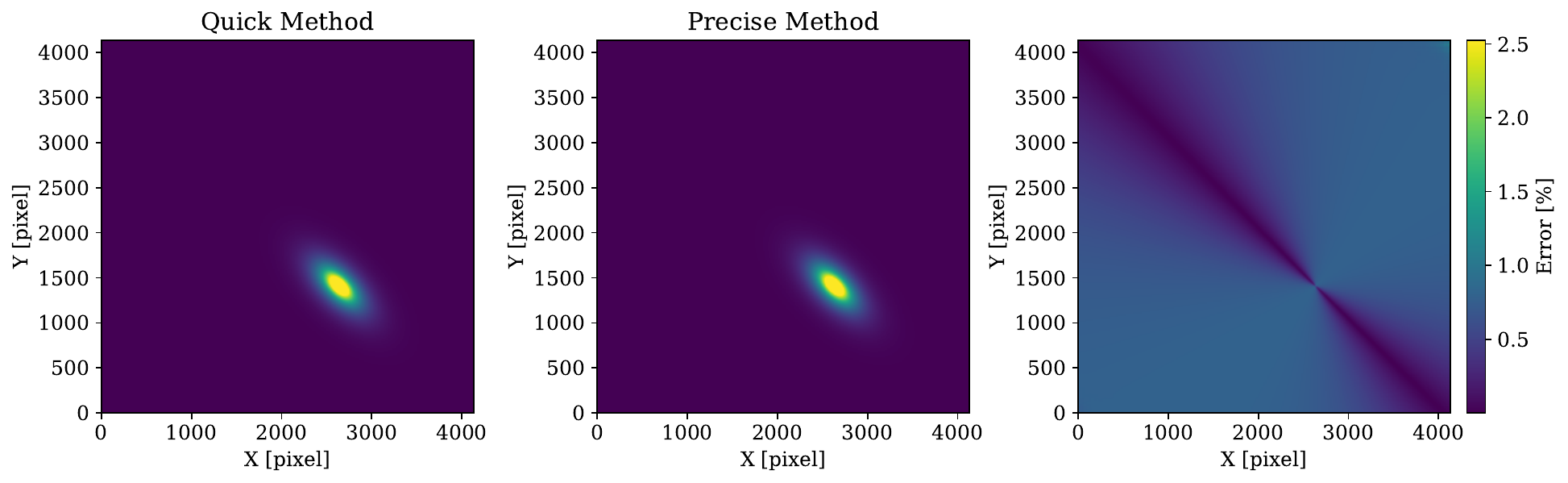}
        \caption{\textit{Left}: Extended source with $n = 1$ created using the {\tt Sersic2D} model within {\tt Astropy}, \textit{Middle}: The same extended source created using the {\tt SersicDistribution} model within \pandeia. \textit{Right}: A measure of the error resulting from using the quick approximation as opposed to the precise method. \label{fig:extended_compare}}
    \end{center}
\end{figure*}

\subsection{Injecting an Extended Source}\label{sec:extended}

In addition to point sources, STIPS has the ability to inject extended sources using two different methods. The first ``quick" method generates extended sources using the {\tt Sersic2D} model within {\tt Astropy} \citep{astropy:2022}. The second ``precise" method uses the {\tt SersicDistribution} profile model within \pandeia to generate the shape of the extended source. Both methods generate a Sérsic profile that takes as input parameters: the center X, Y in pixels of the source, a flux $f$ in Jansky, a Sérsic index $n$, a radius in pixels of $r_e$, an angle in degrees $\phi$, and an axial ratio $R$. The output profile generated is then convolved with the \nancy PSF, generated using the methods outlined in \S\ref{sec:PSFs}, using the {\tt convolve\_fft} function in {\tt Astropy}. In Figure~\ref{fig:extended_compare} we show example scenes created with both methods for a source with $n = 1$, on the right panel we show how the error resulting from using the quick method varies as a function of pixel.

\begin{figure*}
    \begin{center}
        \includegraphics[width=\columnwidth]{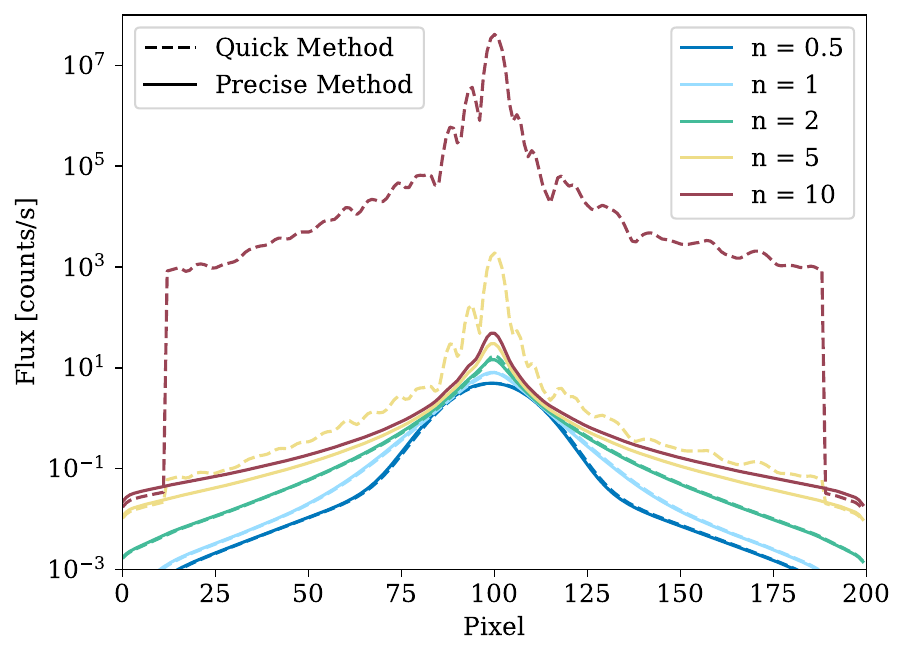}
        \includegraphics[width=\columnwidth]{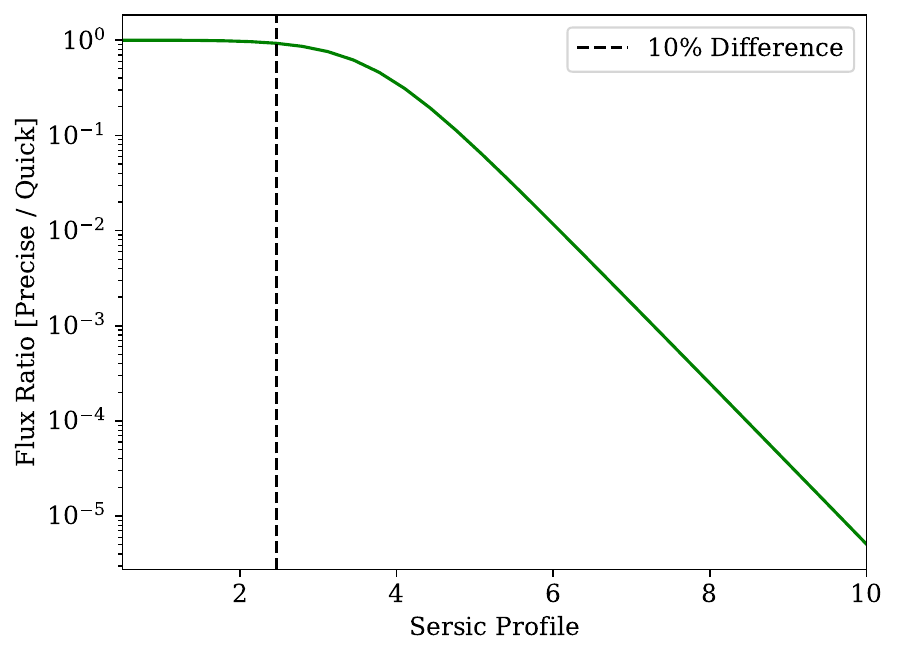}
        \caption{\textit{Left}: Cross-section along the major-axis of a series of extended sources with varying Sérsic profile $n$. The dashed lines are sources created using the quick {\tt Astropy} method, while the solid lines use the precise \pandeia method. \textit{Right}: Ratio of the total integrated flux of an extended source as a function of its Sérsic profile. For $n < 2.5$ the difference between the two methods is $ < 10$\%.
        \label{fig:sersic}}
    \end{center}
\end{figure*}

We find that the quick {\tt Astropy} method is $\sim 8$ times faster than the precise \pandeia method. The difference in accuracy between the two methods is heavily dependent on the Sérsic index $n$. For sources with $n < 2$, the two methods are comparable and only diverge on the order of $\sim 1$\%. On the other hand, the difference between the quick and precise method becomes very significant for $n \gtrsim 5$. The left panel of Figure~\ref{fig:sersic} shows a cross-section along the major-axis of a series of extended sources with varying $n$. It is evident that for high values of $n$ the quick method significantly diverges from the expected profile. On the right panel of Figure~\ref{fig:sersic} we show the total integrated flux of the injected extended source as a function of $n$. For a source with $n = 2.5$, the difference in flux is $\sim 10$\%, the quick method becomes increasingly less reliable for larger values of $n$.

\begin{deluxetable}{ccc}
	\tablecaption{Filter Parameters \label{tab:background}}
	\tablehead{\colhead{Filter} & \colhead{Background} & \colhead{Zeropoint} \\
                   \colhead{} & \colhead{counts / s} & \colhead{AB Mag} }
	\startdata
        F062 & 0.78 & 26.77 \\
        F087 & 0.66 & 26.43 \\
        F106 & 0.62 & 26.45 \\
        F129 & 0.54 & 26.47 \\
        F146 & 1.78 & 27.70 \\
        F158 & 0.49 & 26.50 \\
        F184 & 0.45 & 26.12 \\
        F213 & 3.88 & 26.06 \\
        \enddata
    \tablecomments{Default background and zeropoint  for each WFI filter used by STIPS. Background units are in counts per second}
\end{deluxetable}

The quick method is recommended if speed is the main priority and a high flux accuracy is not required. For example, when thousands of sources need to be generated. The precise method is recommended when either only a few sources are being simulated, or if the sources all have $n \gtrsim 3$.

\subsection{Sources of Background and Noise}\label{sec:background}

STIPS has the option to add a number of sources of background and residual noise. The user can choose to include a constant background. The default value for the background is the benchmark zodiacal background calculated at lat.$= 266.3$ and lon.$=-50$ defined on June 19, 2023, obtained directly from \pandeia. The default background values in counts per second for each \nancy filter are listed in Table~\ref{tab:background}.

In addition to the constant background, STIPS can include residual sources of noise from Poisson noise, readout noise, dark residuals, flat residuals, and cosmic ray residuals.

\begin{enumerate}
    \item Poisson noise is calculated for each pixel simply as $\sqrt{F}$, where $F$ is the flux in counts.
    \item Readout noise is randomly added to each pixel with a mean of 0 and a standard deviation of 12 counts.
    \item Dark current residuals are directly added from a mock dark file included with STIPS, multiplied by
    the exposure time.
    \item Flat residuals are included by multiplying the image by a 2D array with random noise that increases the standard deviation of the noise by $\sim 0.06$\%.
    \item Cosmic ray residuals are added using the method outlined below.
\end{enumerate}

Cosmic rays are added directly to the image as $3 \times 3$ pixel regions with a full width at half maximum (FWHM) of 0.9 pixels. The probability of each pixel being hit is a function of pixel area, exposure time, and cosmic ray rate fixed to 5 hits$/{\rm cm}^2/$s. The brightness of the individual cosmic rays depends on their input energy in electrons. Cosmic rays are injected with two fixed energies of 600 and 5000 electrons.

\section{Validation}\label{sec:validation}

\subsection{Photometry}\label{sec:photometry}

\begin{figure*}
    \begin{center}
        \includegraphics[width=0.9\textwidth]{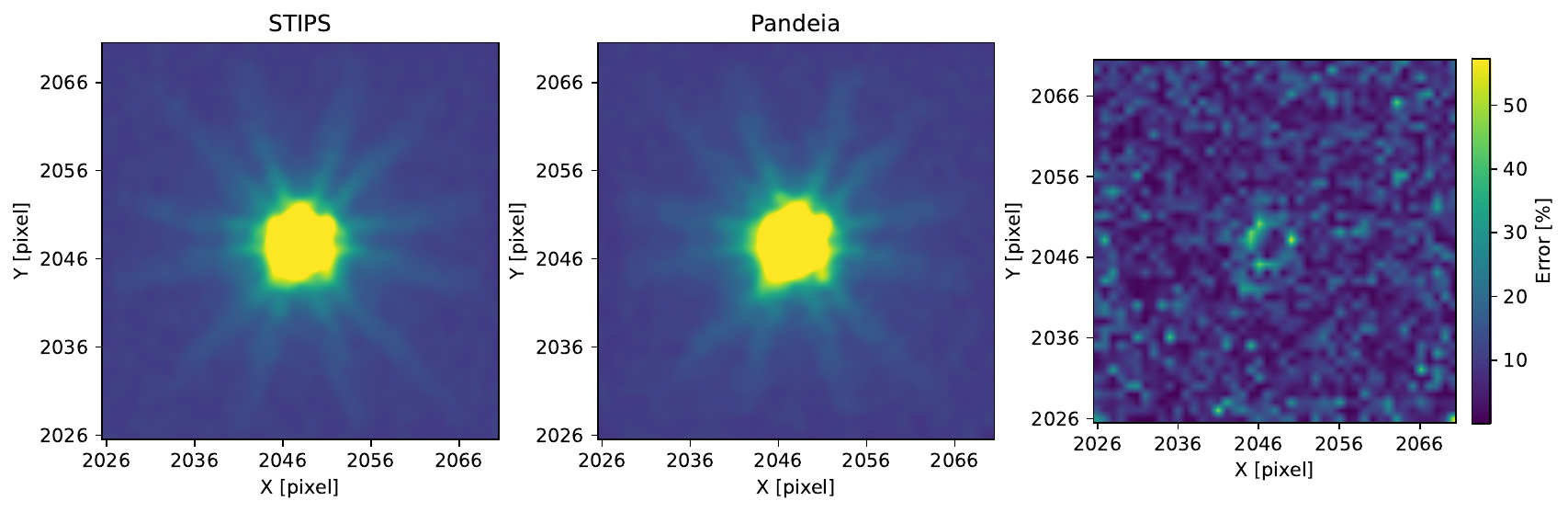}
        \caption{A simulated image of an 18th mag point source in F129 generated with STIPS (\textit{Left}) and with \pandeia (\textit{Middle}). The two results are qualitatively very similar. The \textit{Right} panel shows a map of the error difference between the two PSFs, where the difference is likely dominated by the fact that Pandeia uses an older version of the WFI PSF that does not include distortion effects. \label{fig:pandeia_psf}}
    \end{center}
\end{figure*}

\begin{figure}
    \begin{center}
        \includegraphics[width=\columnwidth]{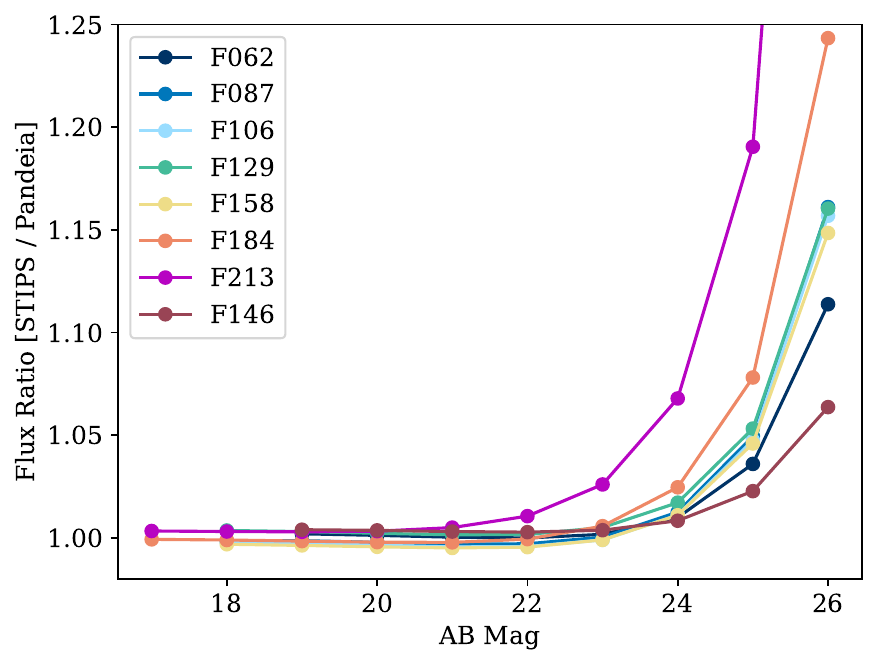}
        \caption{Ratio between the total flux in a simulated STIPS image and the equivalent generated using \pandeia. Saturated sources are excluded from this plot. For bright non-saturated sources, STIPS and \pandeia differ at the $\sim 3$\% level.\label{fig:pandeia_flux}}
    \end{center}
\end{figure}

\begin{figure*}
	\begin{center}
		\includegraphics[width=\textwidth]{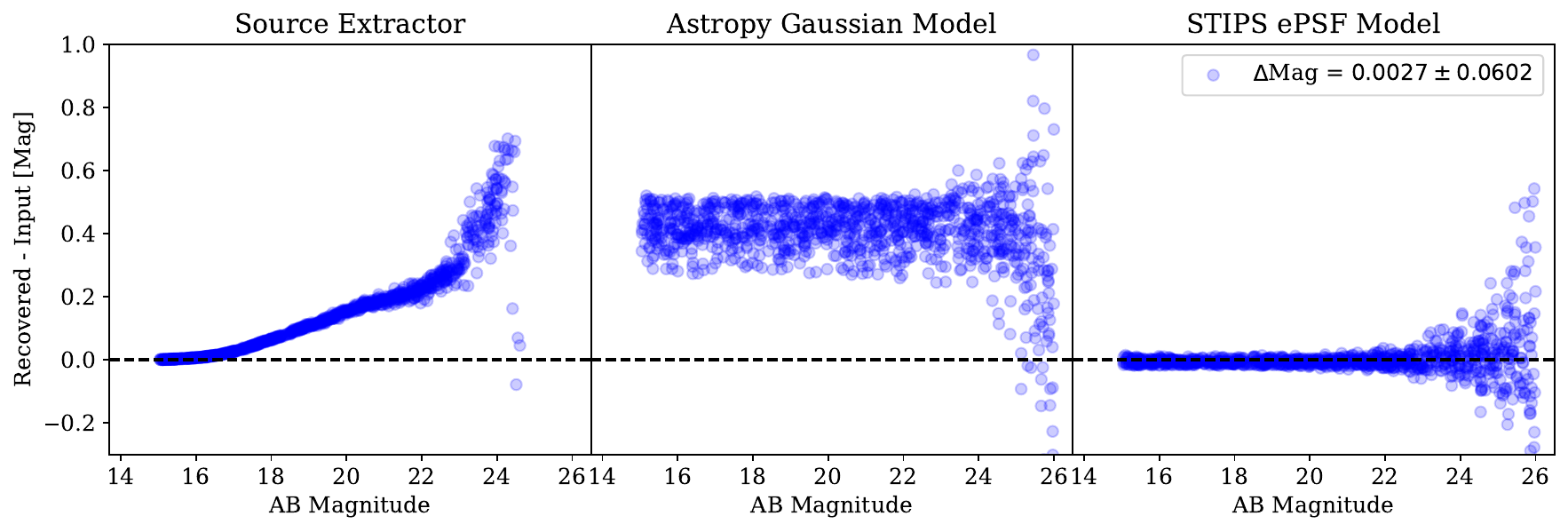}
		\caption{Results from using different tools to recover the magnitudes of 1000 point sources injected into an image with STIPS. \textit{Left}: Source Extractor produces increasingly less accurate results for dimmer sources. \textit{Middle}: Using an {\tt Astropy} Gaussian model results in a clear systematic offset. \textit{Right}: The most accurate method recovers the positions of sources using the same eSPF model that was used to create them in STIPS. \label{fig:photometry}}
	\end{center}
\end{figure*}

We validate the accuracy of the STIPS photometry using two different methods. First, we compare the resulting STIPS simulations to those generated from \pandeia, which we use as a reference benchmark. \pandeia simulations always assume the source is located at a pixel location of (2048, 2048) in SCA1, we therefore fix the STIPS simulations at that position. In Figure~\ref{fig:pandeia_psf} we show a comparison between a STIPS and a \pandeia simulation of a single 18th mag point source in F129. Both images are visually very similar, with minimal deviations between the two away from the core of the ePSF. 90\% of pixels in the image deviate by less then $\sim 10$\%. We note that some of this difference is likely due to the fact that Pandeia uses version 1.0 of {\tt WebbPSF}, while STIPS uses version 1.3, which accounts for detector distortion effects.

In Figure~\ref{fig:pandeia_flux} we show how the total flux of a point source compares between STIPS and \pandeia as a function of magnitude. We calculate the total flux by summing all the counts within the simulated 44 x 44 pixel square PSF after subtracting the mean background in each filter. The STIPS flux values deviate at the level of $\lesssim 1$\% for bright, non-saturated, sources. For sources dimmer than $\sim 25$th mag, the deviation can increase above $\sim 10$\%, depending on the filter. The difference in total flux for dim sources is dominated by the background of the simulations.

The second validation method we use is to inject 1000 point sources between 15th and 26th
magnitude randomly distributed across SCA01 in F129 and then use different methods to do photometry on these sources to recover their magnitudes. For the first method we use the {\tt sep.extract} function within the {\tt Python} implementation of Source Extractor \citep{Bertin96}. The second method uses the {\tt Gaussian2D} model using {\tt Astropy}. The third method fits the same ePSF model created by STIPS to the sources in the image.

Figure~\ref{fig:photometry} shows how the different methods compare. We find that the aperture photometry method in Source Extractor performs well for bright sources, but quickly deviates from the expected magnitude for dimmer sources. While the {\tt Astropy} method performs well and is almost independent of magnitude, the difference in the assumed PSF shape produces a clear systematic offset of $\sim 0.5$ mag from the expected magnitude. Lastly, using the model of the \nancy PSF to fit the simulated sources produces in the most accurate results, where we are able to recover the input magnitudes with a mean deviation of $\sim 0.06$ mag.

\begin{figure*}
	\begin{center}
		\includegraphics[width=\textwidth]{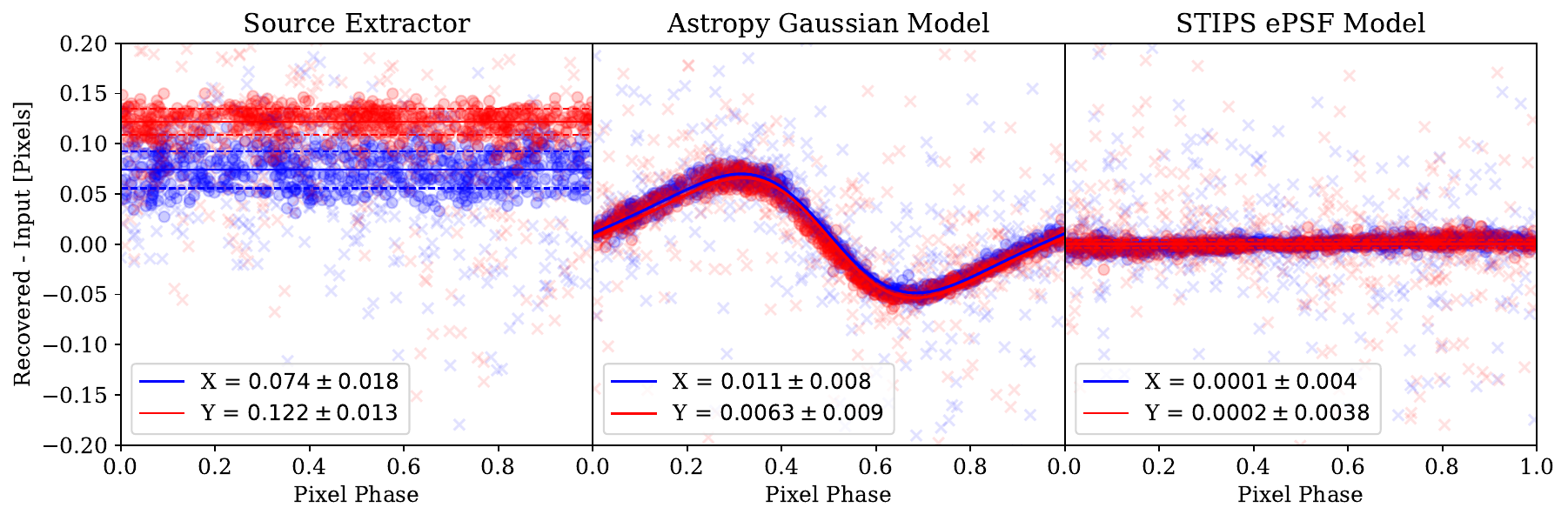}
		\caption{Results from using different tools to recover the positions of 1000 point sources injected into an image with STIPS. The circles represent sources brighter than 21st mag, and the crosses are sources dimmer than this, which have a reduced accuracy. \textit{Left}: Using Source Extractor results in a clear systematic offset due to the asymmetric shape of the \nancy PSF. \textit{Middle}: Using an {\tt Astropy} Gaussian model produces no systematic offset, but still has a clear pixel-phase dependence due to the difference in interpolation method used. \textit{Right}: The most accurate method recovers the positions of sources using the same eSPF model that was used to create them in STIPS. \label{fig:astrometry}}
	\end{center}
\end{figure*}

\subsection{Astrometry}\label{sec:astrometry}

To validate the astrometry of STIPS, we create a simulated scene with 1000 point sources in F129 on SCA1, ranging from 15th to 25th magnitude. We then use the same three tools used to perform the photometry to recover the positions of the sources in the image to make sure their measured output position match their positions in the input catalog. STIPS takes in source coordinates as input in right ascension and declination in degrees, which it then translates into pixel positions using the {\tt astropy} function {\tt wcs\_world2pix}.

The simplest and fastest method we use is the {\tt sep.extract} function within the {\tt Python} implementation of Source Extractor \citep{Bertin96} ran directly on the simulated scene to measure the centers of the sources in the image. The difference between the input and the recovered positions is shown on the left panel of Figure~\ref{fig:astrometry}. We find this method returns a centroid that is systematically offset by $\sim 0.1$ pixels (or $\sim 11$ mas). This is due to the asymmetric shape of the \nancy PSF.

The other method we use is the {\tt Gaussian2D} model within {\tt Astropy}. The difference between the recovered and input positions of sources from using this method are shown in the middle panel of Figure~\ref{fig:astrometry}. For this method, the systematic offset is no longer present, which is made evident by the centroid of the recovered sources being centered around 0. Nevertheless, clear structure is seen as a function of pixel phase. This is a result of the asymmetric PSF being fit with a different interpolation method used within {\tt Astropy} than what STIPS uses. After subtracting this systematic structure, the scatter in the astrometry is $\sim 0.009$ pixels (or $\sim 1$ mas).

Lastly, we use the same ePSF model within STIPS to fit the sources in the image. This method produces the most accurate results, shown in the right panel of Figure~\ref{fig:astrometry}. In this case we are able to recover the input position of sources with an accuracy of $\sim 0.004$ pixels, or $\sim 0.4$ mas, in both the X and Y directions.

\begin{figure}
	\begin{center}
		\includegraphics[width=\columnwidth]{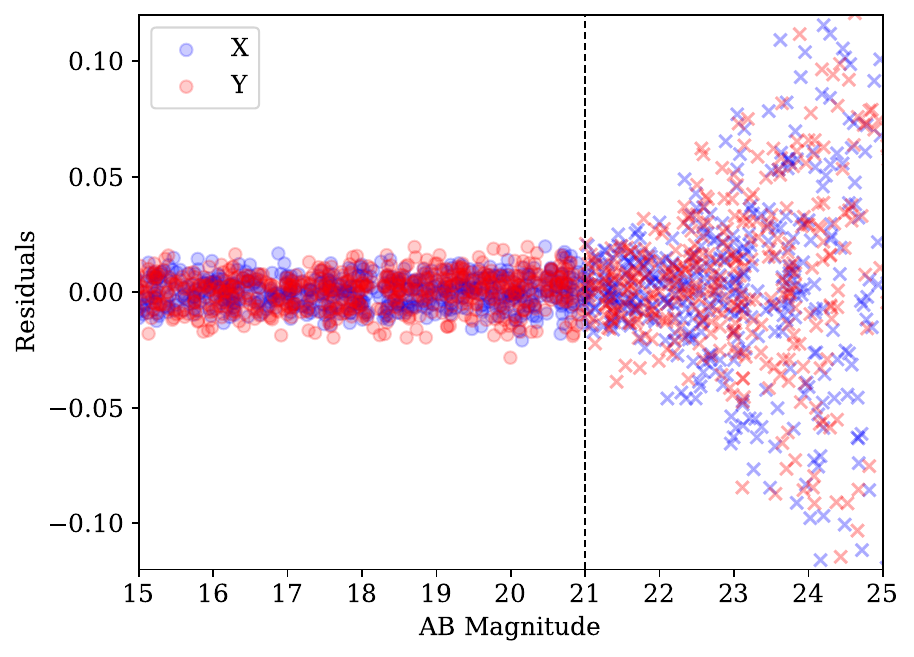}
		\caption{Difference between the recovered and input pixel positions of sources using the STIPS ePSF model, as a function of magnitude. For the specific simulation setup described in \S\ref{sec:astrometry}, the background becomes noticeable for sources dimmer than 21st mag, and we therefore exclude these from our metrics. \label{fig:astrometry_mag}}
	\end{center}
\end{figure}

The accuracy of all methods depends on the magnitude of the source. In Figure~\ref{fig:astrometry_mag} we show the recovered minus input position residuals as a function of source magnitude. For sources dimmer than 21st mag the background begins to have a measurable effect on the accuracy of the centroid measurement.

\section{Usage}\label{sec:usage}

STIPS is designed to be a rapid simulator which creates scenes in units of counts per second, and not a high-accuracy exposure time calculator. Therefore, regardless of the science case at hand, users should take into account the strengths and limitations of STIPS outlined in this paper when deciding which tool to use to simulate \nancy images.

STIPS is well suited for users who wish to create a large number of quick simulations with varying parameters, such as different pointings, filters, background levels, or varying parameters of point sources and extended sources. This functionality allows users to explore how differences in simulation parameters might impact their observations. Additionally, STIPS can be used to simulate subtle variations in the PSF across the WFI, making it useful for testing differences between SCAs.

The STIPS codebase is publicly available on the STScI GitHub\footnote{\url{https://github.com/spacetelescope/STScI-STIPS}}. For the latest instructions on how to install STIPS and examples of how to run the code, we encourage the user to go to the Roman User Documentation\footnote{\url{https://roman-docs.stsci.edu/simulation-tools-handbook-home/stips-space-telescope-imaging-product-simulator}}.

\acknowledgments
S. G. is supported in part by an STScI Postdoctoral Fellowship. We thank Karoline Gilbert, Gisella de Rosa, and Charles-Philippe Lajoie for their contributions on the creation and continued support of this tool. We thank Eddie Schlafly for his insights on romanisim. This research has made use of NASA’s Astrophysics Data System. This research made use of Photutils, an Astropy package for detection and photometry of astronomical sources \citep{Photutils_4624996}. Software citation information aggregated using \texttt{\href{https://www.tomwagg.com/software-citation-station/}{The Software Citation Station}} \citep{software-citation-station-paper, software-citation-station-zenodo}. The authors would like to thank Jeffrey W. Kruk, \nancy project scientist at Goddard Space Flight Center, for his support and guidance regarding reference data. 

\software{\texttt{astropy} \citep{astropy:2013, astropy:2018, astropy:2022}, \texttt{matplotlib} \citep{Hunter2007}, \texttt{numpy} \citep{numpy}, POPPY \citep{Perrin12}, WebbPSF \citep{Perrin14}, synphot \citep{synphot}, Pandeia \citep{Pontoppidan16}, Photutils \citep{Bradley22}, \texttt{python} \citep{python}, \texttt{scipy} \citep{2020SciPy-NMeth, scipy_11702230}, Source Extractor \citep{Bertin96}, and \texttt{scikit-image} \citep{scikit-image}}.

\bibliography{references}

\end{document}